\documentclass[runningheads]{llncs}
\usepackage[T1]{fontenc}
\usepackage{marvosym}
\usepackage{graphicx,verbatim}
\usepackage{booktabs}
\usepackage{graphicx} 
\usepackage{amsmath}
\usepackage{multirow} 
\usepackage{graphicx}
\usepackage{xcolor}
\usepackage[table]{xcolor}
\newcommand{\micon}[1]{\raisebox{-0.15em}{\includegraphics[height=0.8em]{#1}}}
\newcommand{\mname}[1]{\textsf{#1}}

\usepackage{hyperref}
\hypersetup{
    colorlinks=true,
    citecolor=blue,
    linkcolor=red,
    urlcolor=cyan,
}

\begin{document}

\title{AtomiMed: Hierarchical Atomic Fact-Checking for Universal Clinical-Aware Medical Report Evaluation}
\titlerunning{AtomiMed: Atomic Fact-Checking for Medical Report Evaluation}
%
\author{Yuan Wang\inst{1,2}\thanks{Equal contribution. \textsuperscript{\Letter}Corresponding author.} 
\and Wanxing Chang\inst{3,4}$^{*}$  
\and Songtao Jiang\inst{1,2} 
\and Shujian Gao \inst{5} 
\and Xiaotian Zhang\inst{1,2} 
\and Ruifeng Yuan \inst{3,4} 
\and Weiwei Cao \inst{3,4} 
\and Bowen Shi \inst{3,4} \and 
Ling Zhang \inst{3} \and 
Zuozhu Liu \inst{1,2}$^{\href{mailto:zuozhuliu@intl.zju.edu.cn}{\textrm{\Letter}}}$ 
\and Jianpeng Zhang \inst{3,4}$^{\href{mailto:zuozhuliu@intl.zju.edu.cn}{\textrm{\Letter}}}$} 

\authorrunning{Yuan Wang, et al.}

\institute{Zhejiang University-University of Illinois Urbana-Champaign Institute, Zhejiang University, Zhejiang, China
\and College of Computer Science and Technology, Zhejiang University, Zhejiang, China \and
DAMO Academy, Alibaba Group, Zhejiang, China \and
Hupan Lab, Zhejiang, China \and
Institute of Trustworthy Embodied AI, Fudan University, Shanghai, China 
\email{\{yuan2.24\}@intl.zju.edu.cn}\\
\url{https://github.com/Venn2336/MRGEvalkit}}

\maketitle              
\begin{abstract}
Traditional metrics for Medical Report Generation (MRG) predominantly rely on surface-level n-gram overlap, which fails to capture clinical factual accuracy and often overlooks catastrophic diagnostic errors. We address this fundamental limitation by proposing \textbf{AtomiMed}, a universal, modality-agnostic evaluation framework that decomposes complex medical narratives into a standardized, multi-level hierarchy of Atomic Clinical Facts, encompassing Disease-level entities and Attribute-level descriptors, including location, morphology, and severity. By implementing an Agentic Cross-Verification loop between ground-truth and predicted reports, AtomiMed simulates a multi-radiologist peer-review process to verify clinical consistency, thus enabling the decoupled assessment of diagnostic detection and descriptive accuracy. To facilitate standardized evaluation, we introduce \textbf{MRGEvalKit}, an open-source toolkit for automated hierarchical extraction, and curate \textbf{OmniMRG-Bench}, a comprehensive multi-modal benchmark covering X-ray, CT, MRI, and Ultrasound. Extensive experiments on multiple expert-annotated reader studies demonstrate that AtomiMed achieves significantly higher correlation with human radiologist judgment compared to traditional and model-based metrics.


\end{abstract}
\section{Introduction}
Automated Medical Report Generation (MRG) has emerged as a critical capability for alleviating radiologist workload and expanding diagnostic accessibility~\cite{wang2025v2t,moor2023foundation,Wu2023TowardsGF,tu2024towards,wang2025beyond,liu2024medcot,jiang2025omniv}. As MRG systems proliferate, evaluation metrics have become the primary arbiters of system quality; however, the reliable evaluation of generated reports remains an insufficiently studied challenge. A metric that fails to detect clinically critical errors, such as missing a pneumothorax or inverting laterality, provides a false sense of system quality and directly risks patient safety~\cite{oakden2020hidden}. Unlike general NLG, medical report evaluation demands factual grounding, clinical structure awareness, and robustness across diverse imaging modalities~\cite{ostmeier2024green,zhao2024ratescore}.

Existing metrics exhibit systematic and cascading deficiencies. Lexical metrics (BLEU~\cite{papineni2002bleu}, ROUGE~\cite{lin2004rouge}, METEOR~\cite{banerjee2005meteor}) are semantically blind, assigning near-identical scores to no pleural effusion and pleural effusion~\cite{miura2021improving,delbrouck2024radgraph,ostmeier2024green}. Structure-based metrics like CheXbert~\cite{smit2020combining}, RadGraph~\cite{delbrouck2024radgraph,jain2021radgraph}, SembScore~\cite{smit2020combining}, and RaTEScore~\cite{zhao2024ratescore} improve factuality but sacrifice generality. In particular, CheXbert covers only 14 chest X-ray labels, and RadGraph extractors are trained predominantly on chest radiography, precluding their applicability to CT, MRI, etc. LLM-as-a-Judge approaches like GREEN~\cite{ostmeier2024green} achieve stronger radiologist correlation but offer no per-finding audit trace and incur substantial inference costs. Crucially, no existing metric simultaneously addresses modality universality, fine-grained attribute-level correctness, and interpretable error attribution.

We identify the root cause of these failures as a mismatch between holistic report comparison and the inherently hierarchical, compositional structure of clinical narratives: a radiology report is not an atomic document but a structured composition of \textbf{disease-level} presence claims and \textbf{attribute-level} descriptors: location, severity, morphology, among others. Inspired by the peer-review workflow in radiology, where a second reader independently evaluates the same study, we propose decomposing each report into a canonical set of \textbf{Atomic Clinical Facts (ACFs)} and bidirectionally verifying their consistency through an agentic cross-validation loop. This bidirectional design inherently separates diagnostic detection from descriptive accuracy, yielding both an aggregated scalar score and question-level audit traces for interpretable error attribution.

Concretely, we present \textbf{AtomiMed}, a modality-agnostic evaluation framework comprising: (i) a hierarchical atomic decomposition module that extracts Disease-level and Attribute-level QA pairs from any medical report; (ii) an 
Agentic Cross-Verification loop that bidirectionally queries each report as evidence for the other's questions, computing precision, recall, and F1 at both levels; and (iii) \textbf{OmniMRG-Bench}, the first multi-modal benchmark 
spanning X-ray, CT, MRI, and Ultrasound with expert radiologist annotations. Experiments on four reader-study benchmarks demonstrate that AtomiMed achieves significantly higher correlation with radiologist judgment than all prior metrics, including GREEN, while providing interpretable per-finding error attribution.

\noindent Our main contributions are threefold:
\begin{itemize}

    \item \textbf{AtomiMed}: A modality-agnostic evaluation framework that decomposes medical reports into a two-level ACF hierarchy and verifies consistency via an \textbf{Agentic Cross-Verification} loop, enabling decoupled assessment of diagnostic detection and descriptive accuracy with interpretable per-finding error attribution.

    \item \textbf{OmniMRG-Bench \& MRGEvalKit}: The first multi-modal MRG benchmark spanning X-ray, CT, MRI, and Ultrasound with standardized radiologist annotations (Table~\ref{tab:medical_results}), paired with an open-source toolkit for reproducible hierarchical atomic scoring.

    \item \textbf{Empirical Analyses}: Experiments on four radiologist-annotated benchmarks show AtomiMed achieves state-of-the-art expert correlation across modalities, with granular per-finding audit traces exposing systematic attribute- and disease-level deficiencies invisible to all prior metrics.

\end{itemize}

\begin{table}[t!]
    \centering
    \scriptsize
    \setlength{\tabcolsep}{3pt} 
    \renewcommand{\arraystretch}{0.95}
    \caption{MRG performance across multi-modal datasets evaluated by \textbf{AtomiMed}. We report the clinical accuracy scores for both general-purpose vision-language models and medical-specialized models across Radiology (X-Ray, CT, MRI) and Medical (Ultrasound). \textbf{Bold} indicates the best score and \underline{Underline} means the second-best.}
    \label{tab:medical_results}
    \resizebox{\linewidth}{!}{%
        \begin{tabular}{lcccccccc} 
            \toprule
            \textbf{Model} & \multicolumn{3}{c}{\textbf{X-Ray}} & \multicolumn{3}{c}{\textbf{CT}} & \textbf{MRI} & \textbf{Ultrasound} \\
            \cmidrule(lr){2-4} \cmidrule(lr){5-7} \cmidrule(lr){8-8} \cmidrule(lr){9-9}
            & \scriptsize{MIMIC} & \scriptsize{IU-Xray} & \scriptsize{CheXpert} & \scriptsize{CT-RATE} & \scriptsize{AMOS} & \scriptsize{Merlin} & \scriptsize{RadGenome} & \scriptsize{KMVE} \\
            & \scriptsize{(2343)} & \scriptsize{(590)} & \scriptsize{(234)} & \scriptsize{(3039)} & \scriptsize{(400)} & \scriptsize{(5125)} & \scriptsize{(651)} & \scriptsize{(1474)} \\
            \midrule
            \multicolumn{9}{l}{\emph{General Vision-Language Models}} \\
            \midrule
            \micon{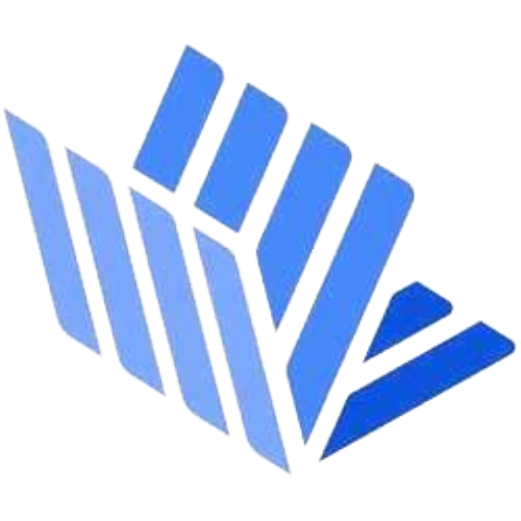}~\mname{InternVL3.5-38B}~\cite{wang2025internvl3} & 0.371 & 0.620 & 0.330 & 0.102 & 0.115 & 0.078 & \textbf{0.527} & 0.281 \\
            \micon{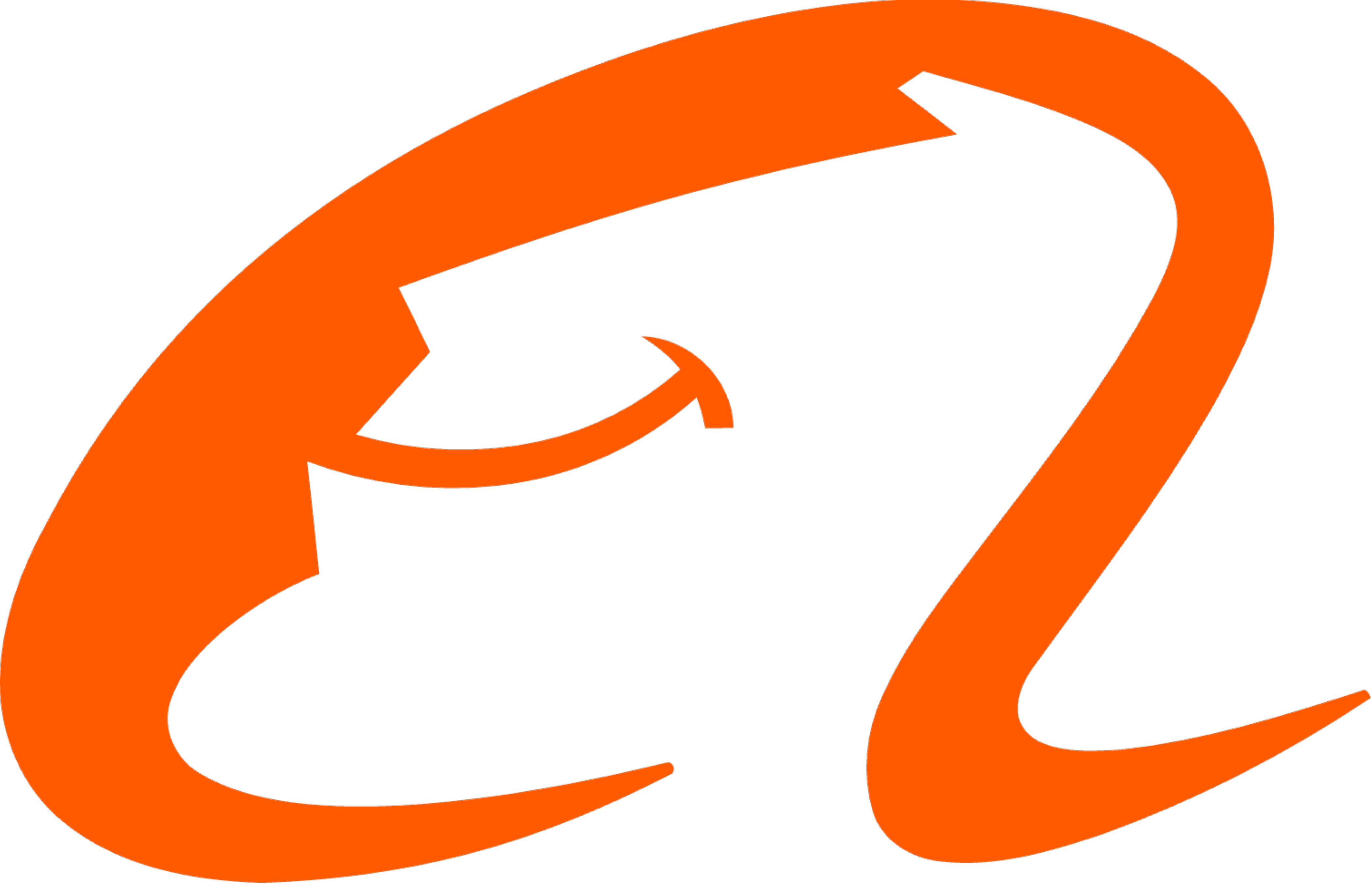}~\mname{Qwen2.5VL-7B}~\cite{bai2025qwen3}    & 0.159 & 0.421 & 0.180 & 0.081 & 0.026 & 0.015 & 0.215 & 0.259 \\
            \micon{qwen.pdf}~\mname{Qwen3VL-8B}~\cite{bai2025qwen3}      & 0.270 & 0.248 & 0.271 & 0.132 & 0.055 & 0.050 & 0.483 & 0.249 \\
            \midrule
            \multicolumn{9}{l}{\emph{Medical-specialized Models}} \\
            \midrule
            \micon{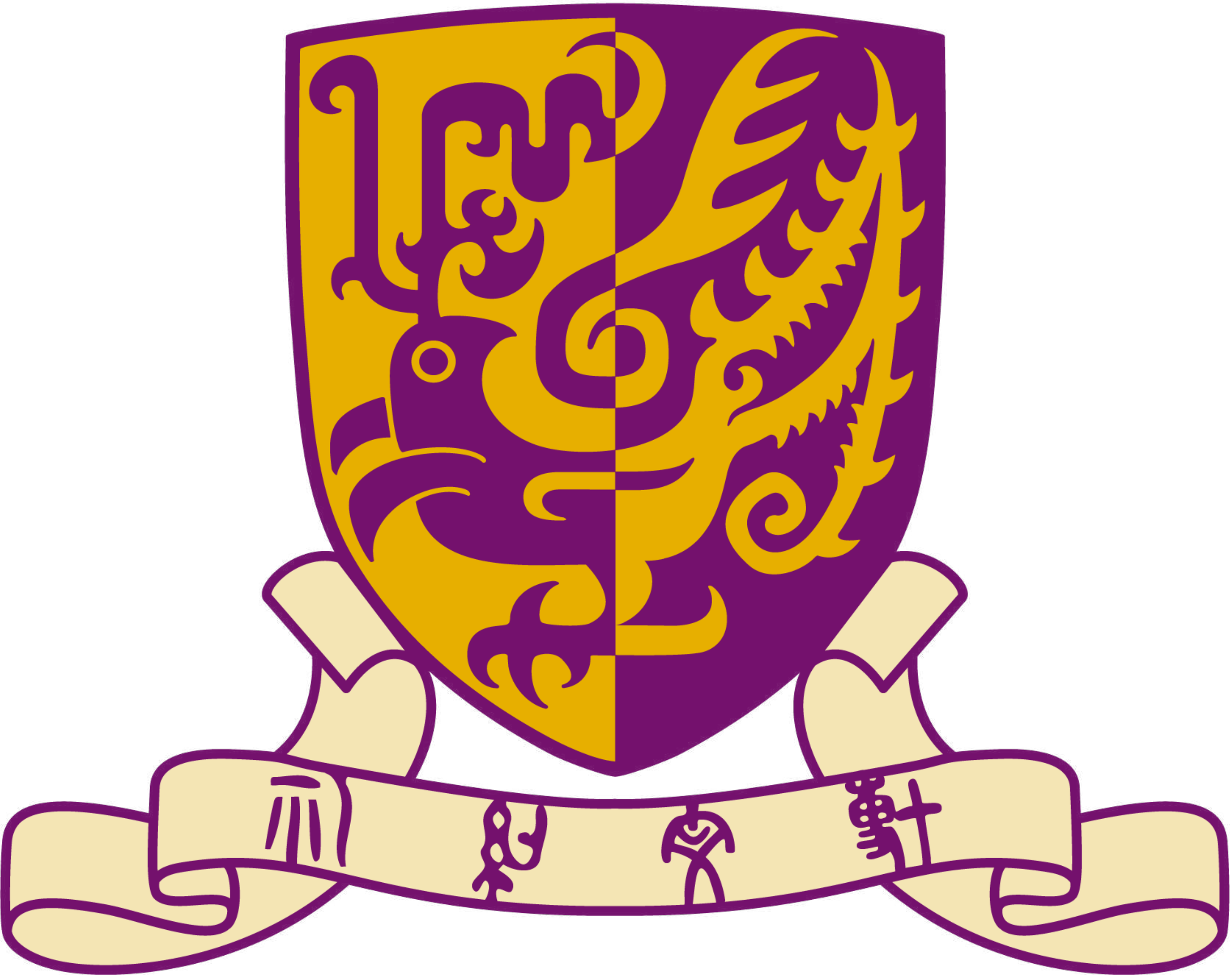}~\mname{HuatuoGPT-34B}~\cite{zhang2023huatuogpt}   & 0.264 & 0.563 & 0.246 & 0.066 & 0.037 & 0.020 & 0.215 & \textbf{0.337} \\
            \micon{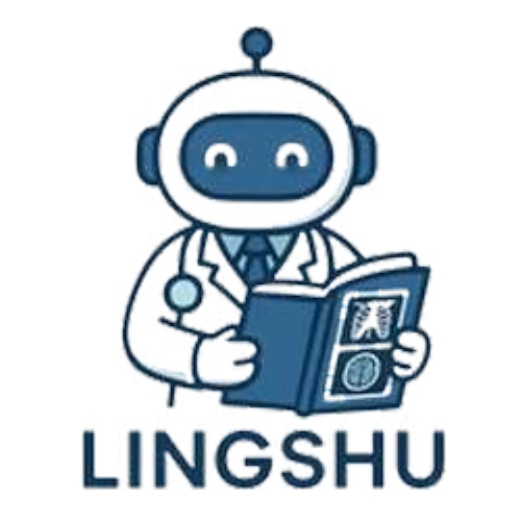}~\mname{Lingshu-7B}~\cite{xu2025lingshu}      & 0.300 & 0.633 & 0.267 & 0.077 & \underline{0.146} & 0.051 & \underline{0.485} & 0.195 \\
            \micon{lingshu_big.pdf}~\mname{Lingshu-32B}~\cite{xu2025lingshu}     & 0.293 & 0.619 & 0.286 & \textbf{0.198} & 0.063 & 0.077 & 0.396 & 0.251 \\
            \micon{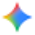}~\mname{MedGemma-27B}~\cite{sellergren2025medgemma}    & 0.328 & 0.524 & 0.266 & 0.083 & 0.056 & 0.056 & 0.429 & 0.158 \\
            \micon{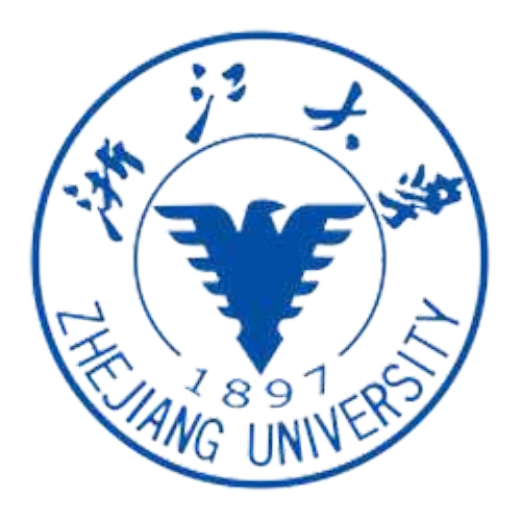}~\mname{HuluMed-7B}~\cite{jiang2025hulu}      & \textbf{0.416} & \underline{0.639} & \underline{0.385} & 0.092 & \underline{0.146} & 0.055 & 0.310 & 0.269 \\
            \micon{hulu_med_big.pdf}~\mname{HuluMed-14B}~\cite{jiang2025hulu}     & \underline{0.405} & \textbf{0.649} & \textbf{0.397} & 0.066 & 0.099 & \underline{0.086} & 0.432 & 0.231 \\
            \micon{hulu_med_big.pdf}~\mname{HuluMed-32B}~\cite{jiang2025hulu}     & 0.395 & 0.637 & 0.383 & \underline{0.145} & \textbf{0.234} & \textbf{0.137} & 0.288 & \underline{0.326} \\
            \bottomrule
        \end{tabular}%
    }
\end{table}

\section{Method}
\subsection{Atomic Decomposition}
\begin{figure*}[htb]
\centering
\includegraphics[width=1.0\textwidth]{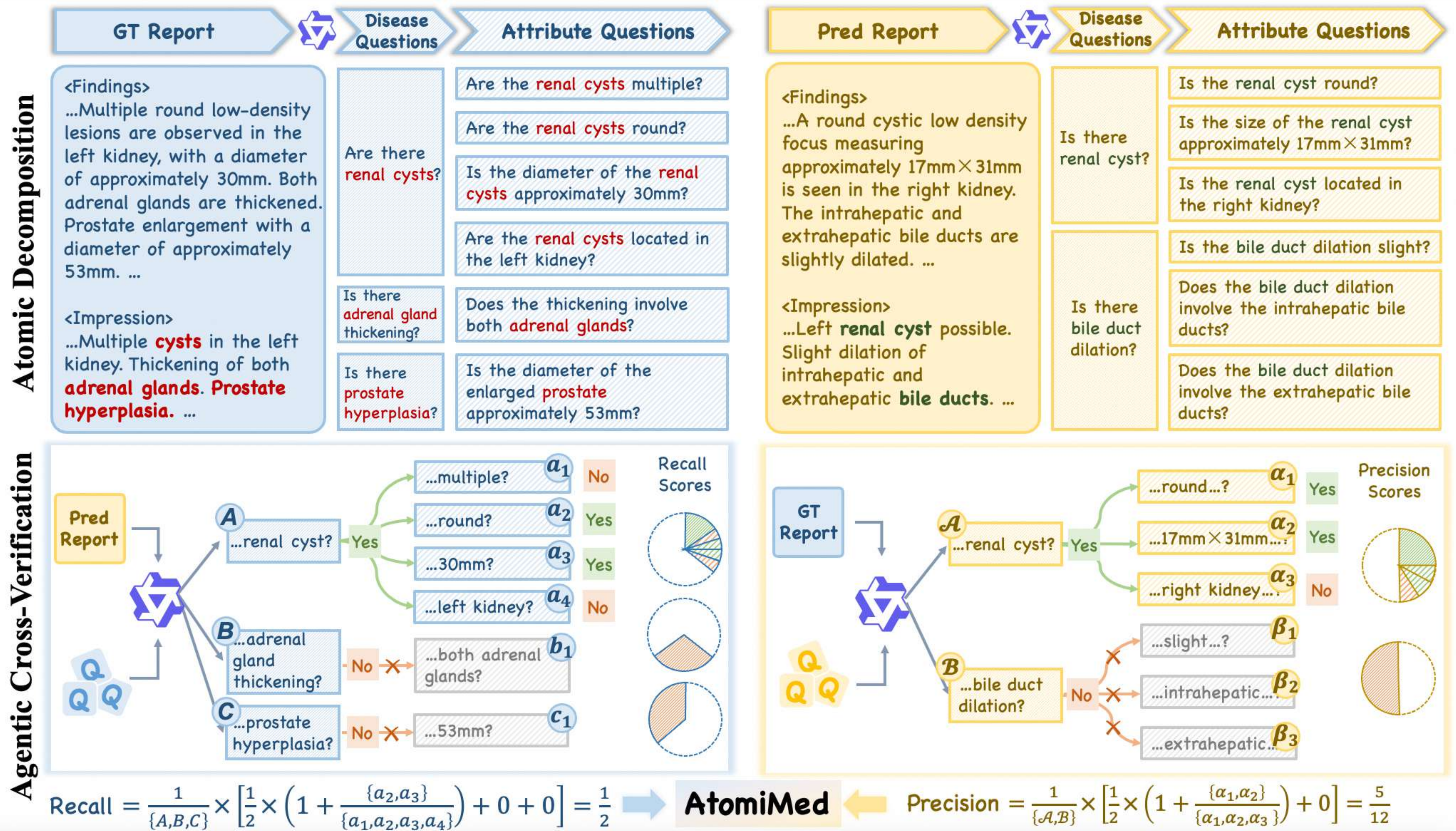}
\caption{\textbf{AtomiMed evaluation framework.} The pipeline consists of two stages: (1) Hierarchical Atomic Decomposition, which extracts Disease-level and Attribute-level QA from reports; and (2) Agentic Cross-Verification, a bidirectional loop that verifies clinical consistency between GT and Pred through evidence-based question answering.}
\label{fig:pipeline}
\end{figure*}

As illustrated in Fig.~\ref{fig:pipeline}, given a report $R$, we prompt an instruction-tuned LLM $\mathcal{M}$ to decompose its clinical content into a two-level hierarchy of ACFs.

\noindent\textbf{Disease-level QA.} The first level captures the presence or absence of each clinical finding as a binary question-answer pair:
\begin{equation}
    \mathcal{Q}^{\mathrm{dis}}(R) = \{(q_i, a_i)\}_{i=1}^{N},
    \quad a_i \in \{\texttt{yes}, \texttt{no}\}
\end{equation}
where each $q_i$ instantiates a normalized clinical entity.

\noindent\textbf{Attribute-level QA.}
The second level associates each identified finding $d_k$ with a set of descriptive facets—location, size, morphology, severity, quantity, and temporal change—formalized as:
\begin{equation}
    \mathcal{Q}^{\mathrm{attr}}(R) = 
    \bigl\{(d_k,\, \{(q_{k,j}, a_{k,j})\}_{j=1}^{M_k})
    \bigr\}_{k=1}^{K}
\end{equation}
This hierarchical decomposition transforms an unstructured narrative into a structured, verifiable ACF set, with $\mathcal{M}$ constrained to emit valid JSON via a fixed prompt template and robust parsing pipeline.
\subsection{Agentic Cross-Validated Assessment}

To quantify clinical consistency between a reference report 
$R_{\mathrm{gt}}$ and a generated report $R_{\mathrm{inf}}$, 
we implement a bidirectional Agentic Cross-Verification
loop that uses $\mathcal{M}$ as an evidence reader.

\noindent\textbf{Disease-level scoring.}
In the GT$\!\to\!$INF direction (recall), each question $q \in \mathcal{Q}^{\mathrm{dis}}(R_{\mathrm{gt}})$ is posed against $R_{\mathrm{inf}}$; in the INF$\!\to\!$GT direction (precision), each question from $R_{\mathrm{inf}}$ is posed against $R_{\mathrm{gt}}$. Precision, recall, and F1 are computed from the resulting match counts:
\begin{equation}
    P_{\mathrm{dis}} = \frac{C_{\mathrm{inf}\to\mathrm{gt}}}{N_{\mathrm{inf}}},\quad
    R_{\mathrm{dis}} = \frac{C_{\mathrm{gt}\to\mathrm{inf}}}{N_{\mathrm{gt}}},\quad
    F1_{\mathrm{dis}} = \frac{2P_{\mathrm{dis}}R_{\mathrm{dis}}}{P_{\mathrm{dis}}+R_{\mathrm{dis}}}
\end{equation}
When both reports yield no disease statements ($N_{\mathrm{gt}}\!=\!N_{\mathrm{inf}}\!=\!0$), scores are set to 1 to avoid penalizing true normal studies.

\noindent\textbf{Attribute-level scoring.}
Attribute verification is \emph{conditioned} on disease-level agreement: only findings correctly aligned in both directions contribute attribute questions. Disease names are extracted from question surface forms and matched via fuzzy string similarity ($\theta\!=\!0.8$) to attribute-level keys. Attribute precision and recall are computed analogously over the aligned finding set.

\noindent\textbf{Final aggregation.}
The two levels are combined via equal-weight averaging:
\begin{equation}
    P = \tfrac{1}{2}P_{\mathrm{dis}} + \tfrac{1}{2}P_{\mathrm{attr}},\quad
    R = \tfrac{1}{2}R_{\mathrm{dis}} + \tfrac{1}{2}R_{\mathrm{attr}},\quad
    \mathrm{F1} = \frac{2PR}{P+R}
\end{equation}
This formulation explicitly separates \emph{diagnostic detection} (disease-level) from \emph{descriptive accuracy} (attribute-level), and every score is traceable to a specific mismatched question, enabling interpretable error attribution.

\begin{figure*}[htb]
\centering
\includegraphics[width=1.0\textwidth]{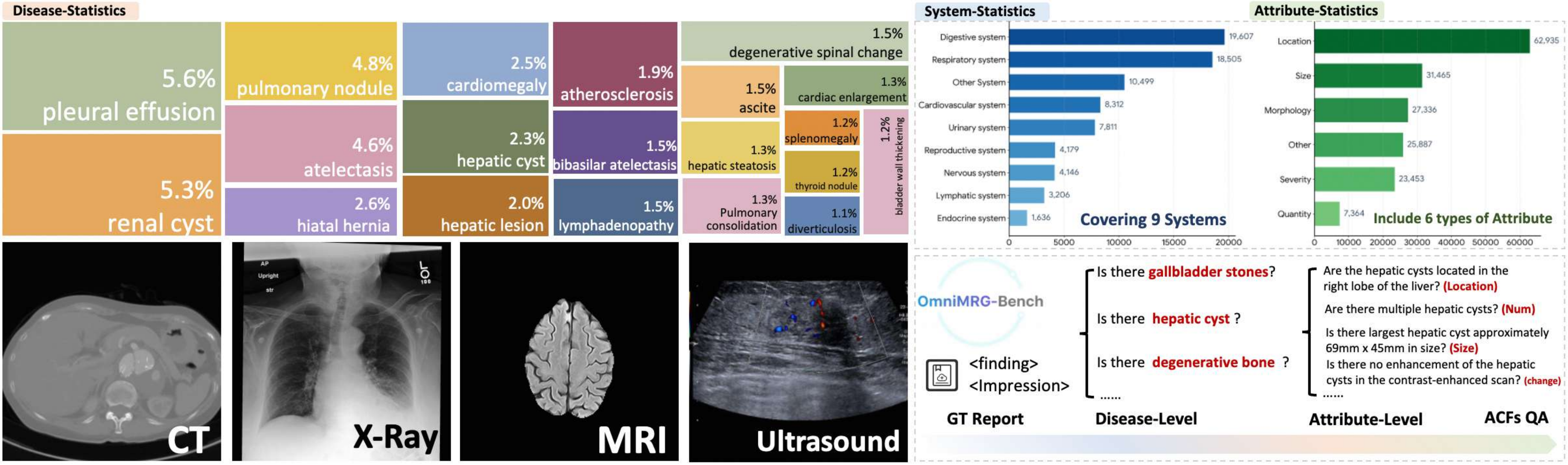}
\caption{Overview of OmniMRG-Bench and MRGEvalKit. This comprehensive multi-modal benchmark spans 9 anatomical systems and 6 attribute categories across X-ray, CT, MRI, and Ultrasound. It comprises over 178K expert-verified, hierarchical ACF pairs to support standardized medical report evaluation.}
\label{fig:omni}
\end{figure*}

\subsection{OmniMRG-Bench}

To support modality-universal evaluation, we curate \textbf{OmniMRG-Bench}, the first multi-modal MRG benchmark spanning four imaging modalities: X-ray, CT, MRI, and Ultrasound. Reports are sourced from publicly available datasets and de-identified clinical archives, covering \textbf{9 anatomical systems} and \textbf{6 attribute categories}, as illustrated in Fig.~\ref{fig:omni}.

ACFs are extracted from ground-truth reports and verified by board-certified radiologists to ensure annotation fidelity. In total, OmniMRG-Bench comprises over 178K disease-level and attribute-level QA pairs, with attribute statistics dominated by location ($\sim$62.9K) and size ($\sim$31.5K) descriptors, enabling broader use as a general-purpose clinical QA benchmark beyond MRG evaluation.

\section{Experiments and Results}

\subsection{Experiment Settings}

\subsubsection{Evaluation Benchmarks and Paradigms.} 
We evaluate \mname{AtomiMed} across two complementary experimental paradigms to rigorously assess both its absolute clinical fidelity and its utility in model selection. 
For \textit{radiologist-correlation analysis}, we utilize four established expert-annotated benchmarks: ReXVal~\cite{yu2023radiology}, containing 600 MIMIC-CXR reports with error counts from six radiologists across categories such as false findings and omissions; ReFiSco-v0~\cite{tian2023refisco}, providing line-level clinical severity annotations; RadEvalX~\cite{calamida2024radiology}, featuring 100 IU-Xray reports with eight distinct error types; and RaTE-Eval~\cite{zhao2024ratescore}, a novel benchmark derived from MIMIC-IV~\cite{johnson2023mimic} and Radiopaedia~\cite{Radiopaedia}. 
To further evaluate clinical-aware preference, we introduce a Dimensionless Pairwise Paradigm. We randomly sample 20 cases each from IU-Xray (X-ray), AMOS (CT), RadGenome (MRI), and KMVE (Ultrasound). For each case, we evaluate the 10 state-of-the-art models listed in Table~\ref{tab:medical_results}, generating a total of $10 \times 10$ model comparison matrix per case to be validated against human expert judgment.

\begin{table}[h]
\centering
\small 
\caption{Alignment with human expert judgment via error count correlation. Kendall’s $\tau$ and Spearman’s $\rho$ correlation coefficients are computed between metric scores and radiologist-annotated error counts across four expert benchmarks. \textbf{Bold} and \underline{underline} denote the best and second-best performance.}
\label{tab:correlation_results}
\resizebox{0.8\textwidth}{!}{%
\begin{tabular}{lcccccccccc}
\toprule
\textbf{Metric} & \multicolumn{2}{c}{\textbf{ReXVal}} & \multicolumn{2}{c}{\textbf{ReFiSco-v0}} & \multicolumn{2}{c}{\textbf{RadEvalX}} & \multicolumn{2}{c}{\textbf{RaTE-Eval}}  \\
 & $\tau$ & $\rho$ & $\tau$ & $\rho$ & $\tau$ & $\rho$ & $\tau$ & $\rho$  \\
\midrule
BLEU-4 & 0.383 & 0.516 & 0.489 & 0.616 & 0.074 & 0.096 & 0.197 & 0.247  \\
ROUGE-L & 0.570 & 0.748 & 0.524 & 0.662 & 0.257 & 0.356 & 0.200 & 0.281  \\
METEOR & 0.484 & 0.653 & 0.468 & 0.617 & 0.201 & 0.284 & 0.174 & 0.245  \\
BERTScore & 0.521 & 0.694 & 0.541 & 0.689 & 0.326 & 0.452 & 0.224 & 0.315  \\
F1 RadGraph & 0.585 & 0.765 & 0.475 & 0.609 & 0.171 & 0.243 & 0.306 & 0.393  \\
SembScore & 0.495 & 0.666 & 0.461 & 0.605 & 0.318 & 0.434 & 0.198 & 0.280  \\
RaTEScore & 0.520 & 0.697 & 0.433 & 0.571 & 0.316 & 0.438 & \underline{0.339} & \textbf{0.460}  \\
GREEN & \underline{0.626} & \underline{0.798} & \underline{0.592} & \underline{0.709} & \underline{0.411} & \underline{0.539} & \textbf{0.374} & \underline{0.457}  \\
Ours & \textbf{0.642} & \textbf{0.806} & \textbf{0.603} & \textbf{0.744} & \textbf{0.432} & \textbf{0.561} & 0.320 & 0.413  \\
\bottomrule
\end{tabular}
}
\end{table}

\subsubsection{Baselines.} 
We compare AtomiMed against a diverse spectrum of metrics: 
(i) \textit{Lexical metrics}, including BLEU-1/4~\cite{papineni2002bleu}, ROUGE-L~\cite{lin2004rouge}, METEOR~\cite{banerjee2005meteor}, and CIDEr~\cite{vedantam2015cider}, representing traditional n-gram overlap; 
(ii) \textit{Embedding-based metrics}, specifically BERTScore~\cite{zhang2019bertscore}, to evaluate semantic similarity via contextual embeddings; 
(iii) \textit{Medical-Specialized metrics}, including F1 RadGraph~\cite{delbrouck2024radgraph,jain2021radgraph}, SembScore~\cite{smit2020combining}, and the graph-based RaTEScore~\cite{zhao2024ratescore}; and 
(iv) \textit{LLM-as-a-Judge}, using the current state-of-the-art GREEN~\cite{ostmeier2024green} for holistic clinical validation.

\subsubsection{Implementation and Evaluation Protocol.} 
AtomiMed is implemented using \texttt{Qwen3-235B-A22B} as the backbone engine for both atomic decomposition and cross-verification. To ensure deterministic and reproducible scoring, we set the decoding temperature to $T=0$. Attribute matching utilizes fuzzy string similarity with a heuristic threshold of $\theta=0.8$. For correlation studies, we report Kendall’s $\tau$ and Spearman’s $\rho$ against radiologist error counts. In the pairwise preference study, a board-certified radiologist independently reviewed the 10 models' outputs for all 80 sampled cases to establish a \textbf{Human Preference Gold Standard}. Metric performance is then quantified by Mean Absolute Error (MAE), Ranking Accuracy (ACC), and Kendall’s $\tau$ between the metric-induced preference matrices and the human-annotated matrix.

\subsection{Main Results}

\subsubsection{Correlation with Radiologist Judgments.} 
As summarized in Table~\ref{tab:correlation_results}, AtomiMed demonstrates superior alignment with expert clinical judgment across diverse benchmarks. On ReXVal, our framework achieves a Spearman’s $\rho$ of \textbf{0.806}, outperforming GREEN (0.798) and significantly exceeding traditional NLP metrics. This high correlation suggests that the bidirectional verification of atomic facts effectively captures the diagnostic errors, such as omissions or laterality shifts, that radiologists penalize most heavily. Notably, on RaTE-Eval, while AtomiMed maintains comparable performance, its primary advantage lies in its ability to provide fine-grained, interpretable audit traces for each clinical finding, a feature absent in holistic LLM judges.

\begin{table}[h]
    \centering
    \scriptsize
    \caption{Dimensionless pairwise preference analysis across imaging modalities. We compare AtomiMed against standard NLP and specialized medical metrics using MAE, Acc, and Kendall’s $\tau$ to measure consistency with radiologist preference rankings.}
    \label{tab:metric_comparison}
    \setlength{\tabcolsep}{2.5pt} 
    \renewcommand{\arraystretch}{1.1}
    \resizebox{\linewidth}{!}{%
        \begin{tabular}{lcccccccccccc}
            \toprule
            \textbf{Metric} & \multicolumn{3}{c}{\textbf{X-Ray}} & \multicolumn{3}{c}{\textbf{CT}} & \multicolumn{3}{c}{\textbf{MRI}} & \multicolumn{3}{c}{\textbf{Ultrasound}} \\
            \cmidrule(lr){2-4} \cmidrule(lr){5-7} \cmidrule(lr){8-10} \cmidrule(lr){11-13}
            & MAE$\downarrow$ & Acc$\uparrow$ & Ken$\uparrow$ & MAE$\downarrow$ & Acc$\uparrow$ & Ken$\uparrow$ & MAE$\downarrow$ & Acc$\uparrow$ & Ken$\uparrow$ & MAE$\downarrow$ & Acc$\uparrow$ & Ken$\uparrow$ \\
            \midrule
            BLEU-1 & 0.4361 & 18.89\% & 0.2816 & 0.4039 & 26.89\% & 0.3422 & 0.3450 & 59.17\% & 0.3358 & 0.5517 & 17.67\% & 0.2063 \\
            BLEU-2 & 0.4299 & 19.31\% & 0.3156 & 0.3983 & 27.44\% & 0.3577 & 0.3450 & 59.17\% & 0.3277 & 0.5383 & 19.00\% & 0.1560 \\
            BLEU-3 & 0.4465 & 17.64\% & 0.2380 & 0.4028 & 27.00\% & 0.3295 & 0.3150 & 62.17\% & 0.3911 & 0.5183 & 21.00\% & 0.0639 \\
            BLEU-4 & 0.4535 & 16.94\% & 0.1987 & 0.4017 & 27.11\% & 0.3313 & 0.3250 & 61.17\% & 0.3699 & 0.5050 & 22.33\% & 0.0115 \\
            \midrule
            Rouge-1 & 0.4139 & 20.97\% & 0.3946 & 0.3861 & 28.67\% & 0.4060 & 0.3533 & 58.50\% & 0.3377 & 0.4783 & 25.00\% & 0.0674 \\
            Rouge-2 & 0.4215 & 20.69\% & 0.3269 & 0.4083 & 26.44\% & 0.3046 & 0.3700 & 56.83\% & 0.2771 & 0.4433 & 28.67\% & 0.2097 \\
            Rouge-L & 0.4194 & 20.42\% & 0.3668 & 0.3817 & 29.22\% & 0.4211 & 0.3833 & 55.50\% & 0.2655 & 0.4750 & 25.33\% & 0.0794 \\
            METEOR  & 0.4917 & 13.06\% & 0.0455 & 0.3911 & 27.11\% & 0.4363 & 0.3500 & 58.83\% & 0.3216 & 0.4817 & 24.67\% & 0.0452 \\
            \midrule
            CIDEr   & 0.4229 & 20.00\% & 0.3524 & 0.4128 & 27.22\% & 0.1225 & 0.4600 & 47.83\% & 0.0709 & 0.5683 & 16.00\% & 0.2563 \\
            \midrule
            RaTEScore    & 0.4194 & 20.42\% & 0.3743 & 0.3883 & 28.44\% & 0.3998 & 0.3233 & 61.50\% & 0.3967 & 0.3783 & 35.00\% & \textbf{0.4516} \\
            GREEN   & 0.1857 & 63.57\% & 0.6481 & 0.2964 & 47.14\% & 0.3283 & 0.3792 & 48.75\% & 0.1513 & 0.3950 & 33.83\% & 0.1667 \\
            \rowcolor{gray!10} \textbf{AtomiMed} & \textbf{0.0214} & \textbf{95.71\%} & \textbf{0.9807} & \textbf{0.0917} & \textbf{84.33\%} & \textbf{0.7403} & \textbf{0.2806} & \textbf{68.19\%} & \textbf{0.4500} & \textbf{0.2944} & \textbf{49.86\%} & 0.3889 \\
            \bottomrule
            \multicolumn{13}{l}{\footnotesize $\downarrow$: lower is better, $\uparrow$: higher is better} \\
        \end{tabular}%
    }
\end{table}

\subsubsection{Pairwise Preference and Clinical Awareness.} 
\begin{figure*}[htb]
\centering
\includegraphics[width=0.95\textwidth]{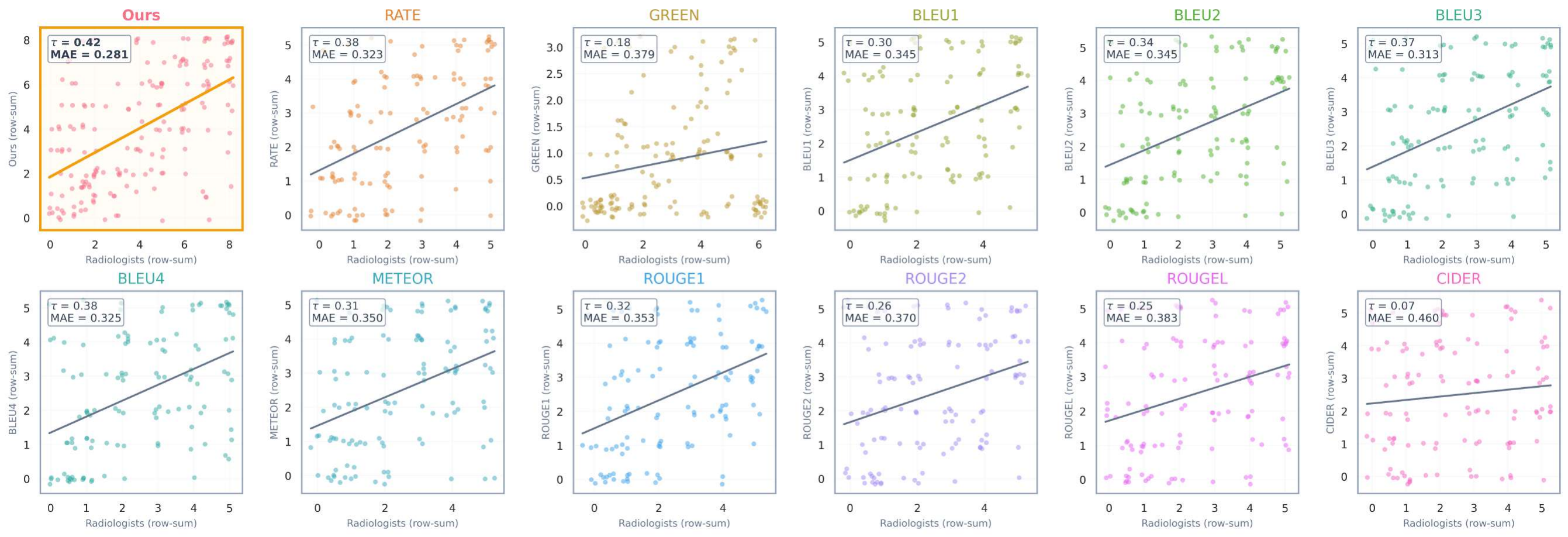}
\caption{Scatter plots of metric scores v.s. human radiologist rankings in MRI.}
\label{fig:Scatter}
\end{figure*}

Table~\ref{tab:metric_comparison} and Fig.~\ref{fig:Scatter} together 
reveal a clear stratification among metrics in their ability to emulate 
radiologist preference.
AtomiMed achieves \textbf{95.71\%} ACC and $\tau\!=\!0.9807$ on X-ray, 
with an MAE of just 0.0214, which is an order of magnitude lower than GREEN 
(MAE~0.1857, ACC~63.57\%) and over twenty times lower than any lexical 
baseline, which uniformly stagnate between 13\% and 21\% ACC regardless 
of modality.
Crucially, GREEN's correlation collapses outside its chest-centric 
training domain: its Kendall's $\tau$ falls from 0.6481 on X-ray to 
0.3283 on CT and 0.1513 on MRI, indicating near-random agreement with 
radiologist preference on cross-sectional imaging. 
This degradation is directly visible in Fig.~\ref{fig:Scatter}, where 
GREEN's scatter in MRI is wide and poorly fitted ($\tau\!=\!0.18$, 
MAE~$=\!0.379$), whereas AtomiMed's points cluster tightly around the 
regression line ($\tau\!=\!0.42$, MAE~$=\!0.281$). 
AtomiMed sustains meaningful correlation across all modalities, reaching 
\textbf{84.33\%} ACC on CT and \textbf{49.86\%} on Ultrasound, where 
GREEN drops to 33.83\% and RaTEScore, the strongest specialist baseline, 
reaches only 35.00\%.

\subsubsection{Granular Performance Analysis via AtomiMed.} 

\begin{figure*}[htbp]
\centering
\includegraphics[width=0.9\textwidth]{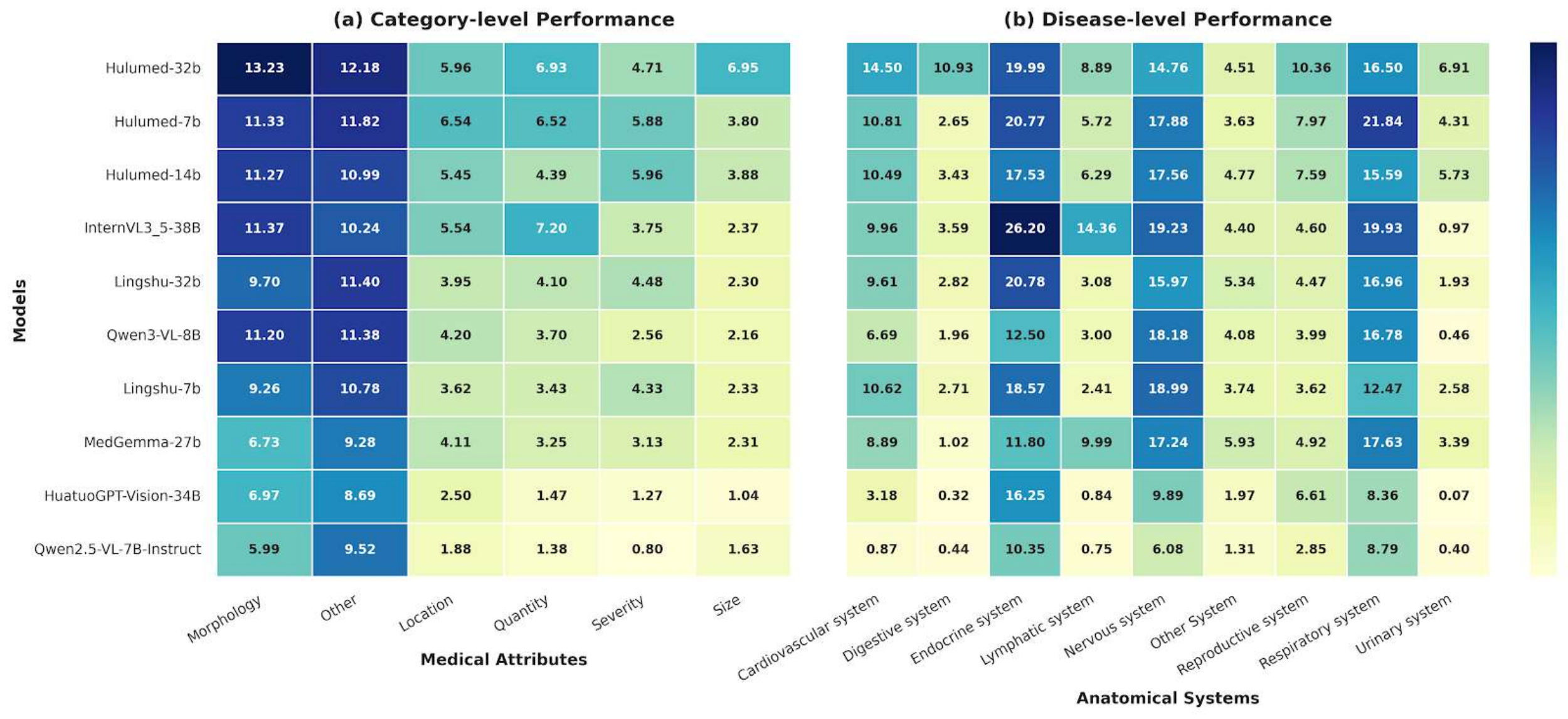}
\caption{Granular performance analysis of models via AtomiMed. Heatmaps illustrating: (a) Category-level performance across medical attributes; and (b) Disease-level performance across various anatomical systems. Higher scores indicate better alignment with human-verified atomic facts.}
\label{fig:analysis}
\end{figure*}

Fig.~\ref{fig:analysis} reveals systematic, non-uniform failure modes that holistic scalar metrics cannot expose, an interpretive capability unique to AtomiMed.
At the \textbf{attribute level} (Fig.~\ref{fig:analysis}a), all models score markedly higher on Morphology (6.0–13.2) yet degrade sharply on Severity (1.3–5.9) and Size (1.0–6.9), with HuatuoGPT-34B (1.27) and Qwen2.5-VL-7B (0.80) near floor on Severity, suggesting models can describe findings qualitatively but consistently fail to quantify clinical significance or precise extent.
At the \textbf{disease level} (Fig.~\ref{fig:analysis}b), the respiratory system dominates (up to 21.84 for HuluMed-7B), reflecting chest-centric pretraining biases, while Digestive (0.32–10.93), Reproductive (1.31–5.34), and Urinary (0.07–6.91) systems are severely underserved. InternVL3.5-38B further exemplifies uneven anatomical coverage, spiking on Endocrine (26.20) yet collapsing on Urinary (0.97), fine-grained deficiencies invisible to any prior evaluation metric.

\section{Conclusion}

We presented AtomiMed, a modality-agnostic framework for MRG evaluation. By decomposing reports into Disease- and Attribute-level Atomic Clinical Facts, AtomiMed utilizes a bidirectional Agentic Cross-Verification loop to operationalize radiological peer review as a computational protocol. Future work will pursue efficient distilled backbone models to reduce inference cost, extend the attribute hierarchy to longitudinal imaging comparisons, and broaden benchmark coverage to additional clinical specialties.

\begin{credits}
\subsubsection{\ackname} The work was done during Yuan’s internship at DAMO Academy. This work is supported by the "Pioneer" and "Leading Goose" R\&D Program of Zhejiang (Grant no. 2025C01128), and the ZJU-Angelalign R\&D Center for Intelligence Healthcare.

\subsubsection{\discintname}
The authors have no competing interests to declare that are relevant to the content of this article.
\end{credits}



%
%
%

\begin{thebibliography}{10}
\providecommand{\url}[1]{\texttt{#1}}
\providecommand{\urlprefix}{URL }
\providecommand{\doi}[1]{https://doi.org/#1}

\bibitem{Radiopaedia}
{Radiopaedia.org}. \url{https://radiopaedia.org}. Accessed: May 2023

\bibitem{bai2025qwen3}
Bai, S., Cai, Y., Chen, R., Chen, K., Chen, X., Cheng, Z., Deng, L., Ding, W., Gao, C., Ge, C., et~al.: Qwen3-vl technical report. arXiv preprint arXiv:2511.21631  (2025)

\bibitem{banerjee2005meteor}
Banerjee, S., Lavie, A.: Meteor: An automatic metric for mt evaluation with improved correlation with human judgments. In: Proceedings of the Acl Workshop on Intrinsic and Extrinsic Evaluation Measures for Machine Translation and/or Summarization. pp. 65--72 (2005)

\bibitem{calamida2024radiology}
Calamida, A.R., Nooralahzadeh, F., Rohanian, M., Nishio, M., Fujimoto, K., Krauthammer, M.: Radiology report generation models evaluation dataset for chest x-rays (radevalx). PhysioNet  (2024)

\bibitem{delbrouck2024radgraph}
Delbrouck, J.B., Chambon, P., Chen, Z., Varma, M., Johnston, A., Blankemeier, L., Van~Veen, D., Bui, T., Truong, S., Langlotz, C.: Radgraph-xl: A large-scale expert-annotated dataset for entity and relation extraction from radiology reports. In: Findings of the Association for Computational Linguistics: ACL 2024. pp. 12902--12915 (2024)

\bibitem{jain2021radgraph}
Jain, S., Agrawal, A., Saporta, A., Truong, S.Q., Duong, D.N., Bui, T., Chambon, P., Zhang, Y., Lungren, M.P., Ng, A.Y., et~al.: Radgraph: Extracting clinical entities and relations from radiology reports. arXiv preprint arXiv:2106.14463  (2021)

\bibitem{jiang2025hulu}
Jiang, S., Wang, Y., Song, S., Hu, T., Zhou, C., Pu, B., Zhang, Y., Yang, Z., Feng, Y., Zhou, J.T., et~al.: Hulu-med: A transparent generalist model towards holistic medical vision-language understanding. arXiv preprint arXiv:2510.08668  (2025)

\bibitem{jiang2025omniv}
Jiang, S., Wang, Y., Song, S., Zhang, Y., Meng, Z., Lei, B., Wu, J., Sun, J., Liu, Z.: Omniv-med: Scaling medical vision-language model for universal visual understanding. arXiv preprint arXiv:2504.14692  (2025)

\bibitem{johnson2023mimic}
Johnson, A.E., Bulgarelli, L., Shen, L., Gayles, A., Shammout, A., Horng, S., Pollard, T.J., Hao, S., Moody, B., Gow, B., et~al.: Mimic-iv, a freely accessible electronic health record dataset. Scientific data  \textbf{10}(1), ~1 (2023)

\bibitem{lin2004rouge}
Lin, C.Y.: Rouge: A package for automatic evaluation of summaries. In: Text Summarization Branches Out. pp. 74--81 (2004)

\bibitem{liu2024medcot}
Liu, J., Wang, Y., Du, J., Zhou, J.T., Liu, Z.: Medcot: Medical chain of thought via hierarchical expert. In: Proceedings of the 2024 Conference on Empirical Methods in Natural Language Processing. pp. 17371--17389 (2024)

\bibitem{miura2021improving}
Miura, Y., Zhang, Y., Tsai, E., Langlotz, C., Jurafsky, D.: Improving factual completeness and consistency of image-to-text radiology report generation. In: Proceedings of the 2021 Conference of the North American Chapter of the Association for Computational Linguistics: Human Language Technologies. pp. 5288--5304 (2021)

\bibitem{moor2023foundation}
Moor, M., Banerjee, O., Abad, Z.S.H., Krumholz, H.M., Leskovec, J., Topol, E.J., Rajpurkar, P.: Foundation models for generalist medical artificial intelligence. Nature  \textbf{616}(7956),  259--265 (2023)

\bibitem{oakden2020hidden}
Oakden-Rayner, L., Dunnmon, J., Carneiro, G., R{\'e}, C.: Hidden stratification causes clinically meaningful failures in machine learning for medical imaging. In: Proceedings of the ACM conference on health, inference, and learning. pp. 151--159 (2020)

\bibitem{ostmeier2024green}
Ostmeier, S., Xu, J., Chen, Z., Varma, M., Blankemeier, L., Bluethgen, C., Md, A.E.M., Moseley, M., Langlotz, C., Chaudhari, A.S., et~al.: Green: Generative radiology report evaluation and error notation. In: Findings of the association for computational linguistics: EMNLP 2024. pp. 374--390 (2024)

\bibitem{papineni2002bleu}
Papineni, K., Roukos, S., Ward, T., Zhu, W.J.: Bleu: a method for automatic evaluation of machine translation. In: Proceedings of the 40th annual meeting of the Association for Computational Linguistics. pp. 311--318 (2002)

\bibitem{sellergren2025medgemma}
Sellergren, A., Kazemzadeh, S., Jaroensri, T., Kiraly, A., Traverse, M., Kohlberger, T., Xu, S., Jamil, F., Hughes, C., Lau, C., et~al.: Medgemma technical report. arXiv preprint arXiv:2507.05201  (2025)

\bibitem{smit2020combining}
Smit, A., Jain, S., Rajpurkar, P., Pareek, A., Ng, A.Y., Lungren, M.: Combining automatic labelers and expert annotations for accurate radiology report labeling using bert. In: Proceedings of the 2020 conference on empirical methods in natural language processing (EMNLP). pp. 1500--1519 (2020)

\bibitem{tian2023refisco}
Tian, K., Hartung, S.J., Li, A.A., Jeong, J., Behzadi, F., Calle-Toro, J., Adithan, S., Pohlen, M., Osayande, D., Rajpurkar, P.: Refisco: Report fix and score dataset for radiology report generation. PhysioNet  (2023)

\bibitem{tu2024towards}
Tu, T., Azizi, S., Driess, D., Schaekermann, M., Amin, M., Chang, P.C., Carroll, A., Lau, C., Tanno, R., Ktena, I., et~al.: Towards generalist biomedical ai. NEJM AI  \textbf{1}(3),  AIoa2300138 (2024)

\bibitem{vedantam2015cider}
Vedantam, R., Lawrence~Zitnick, C., Parikh, D.: Cider: Consensus-based image description evaluation. In: Proceedings of the IEEE conference on computer vision and pattern recognition (CVPR). pp. 4566--4575 (2015)

\bibitem{wang2025internvl3}
Wang, W., Gao, Z., Gu, L., Pu, H., Cui, L., Wei, X., Liu, Z., Jing, L., Ye, S., Shao, J., et~al.: Internvl3. 5: Advancing open-source multimodal models in versatility, reasoning, and efficiency. arXiv preprint arXiv:2508.18265  (2025)

\bibitem{wang2025beyond}
Wang, Y., Gao, S., Liu, J., Jiang, S., Xia, H., Zhang, X., Kang, Z., Wang, Y., Liu, Z.: Beyond n-grams: A hierarchical reward learning framework for clinically-aware medical report generation. arXiv preprint arXiv:2512.02710  (2025)

\bibitem{wang2025v2t}
Wang, Y., Liu, J., Gao, S., Feng, B., Tang, Z., Gai, X., Wu, J., Liu, Z.: V2t-cot: From vision to text chain-of-thought for medical reasoning and diagnosis. In: International Conference on Medical Image Computing and Computer-Assisted Intervention. pp. 658--668. Springer (2025)

\bibitem{Wu2023TowardsGF}
Wu, C., Zhang, X., Zhang, Y., Wang, Y., Xie, W.: Towards generalist foundation model for radiology by leveraging web-scale 2d\&3d medical data. arXiv preprint arXiv:2308.02463  (2023)

\bibitem{xu2025lingshu}
Xu, W., Chan, H.P., Li, L., Aljunied, M., Yuan, R., Wang, J., Xiao, C., Chen, G., Liu, C., Li, Z., et~al.: Lingshu: A generalist foundation model for unified multimodal medical understanding and reasoning. arXiv preprint arXiv:2506.07044  (2025)

\bibitem{yu2023radiology}
Yu, F., Endo, M., Krishnan, R., Pan, I., Tsai, A., Reis, E.P., Fonseca, E., Lee, H., Shakeri, Z., Ng, A., et~al.: Radiology report expert evaluation (rexval) dataset (2023)

\bibitem{zhang2023huatuogpt}
Zhang, H., Chen, J., Jiang, F., Yu, F., Chen, Z., Chen, G., Li, J., Wu, X., Zhiyi, Z., Xiao, Q., et~al.: Huatuogpt, towards taming language model to be a doctor. In: Findings of the association for computational linguistics: EMNLP 2023. pp. 10859--10885 (2023)

\bibitem{zhang2019bertscore}
Zhang, T., Kishore, V., Wu, F., Weinberger, K.Q., Artzi, Y.: Bertscore: Evaluating text generation with bert. In: International Conference on Learning Representations (2020)

\bibitem{zhao2024ratescore}
Zhao, W., Wu, C., Zhang, X., Zhang, Y., Wang, Y., Xie, W.: Ratescore: A metric for radiology report generation. In: Proceedings of the 2024 Conference on Empirical Methods in Natural Language Processing. pp. 15004--15019 (2024)

\end{thebibliography}
%

\end{document}